\documentclass[11pt]{article}
\usepackage{graphicx}
\usepackage[]{amsmath,amssymb}
\usepackage{graphics,epsfig}
\usepackage{amsfonts}
\usepackage{epstopdf}
\def\text#1{\mbox{\rm #1\ }}

\def\ie{{\rm i.e.,\/}\ }
\def\etc{{\rm etc.\/}\ }

\def\one{\mbox{\rm 1}\hskip-2.8pt \mbox{\rm l}}
\def\0 {{\bf 0}}

\newcommand{\CC}{\mathbb{C}}

\def\sqface#1#2#3#4#5#6#7{\rule[-2.8\unitlength]{0in}{5.6\unitlength}
\begin{picture}(4,4)(-#6,-#7)
\put(0,-2){\vector(1,0){4}}
\put(4,2){\vector(0,-1){4}}
\put(0,2){\vector(1,0){4}}
\put(0,2){\vector(0,-1){4}}
\put(0,-3.2){\makebox(0,0)[b]{\scriptsize \mbox{$#1$}}}
\put(4,-3.2){\makebox(0,0)[b]{\scriptsize \mbox{$#2$}}}
\put(4,2.5){\makebox(0,0)[b]{\scriptsize \mbox{$#3$}}}
\put(0,2.5){\makebox(0,0)[b]{\scriptsize \mbox{$#4$}}}
\put(2,0){\makebox(0,0){\scriptsize \mbox{$#5$}}}
\end{picture}}

\def\punit#1{\hspace{#1\unitlength}}

\def\vertgraph#1#2#3#4#5#6#7{\rule[-2.8\unitlength]{0in}{5.6\unitlength}
\begin{picture}(4,4)(-#6,-#7)
\put(0,-2){\vector(1,0){4}}
\put(2,2){\vector(0,-1){4}}
\put(0,2){\vector(1,0){4}}
\put(0,-3.2){\makebox(0,0)[b]{\scriptsize \mbox{$#3$}}}
\put(4,-3.2){\makebox(0,0)[b]{\scriptsize \mbox{$#4$}}}
\put(4,2.5){\makebox(0,0)[b]{\scriptsize \mbox{$#2$}}}
\put(0,2.5){\makebox(0,0)[b]{\scriptsize \mbox{$#1$}}}
\put(2.5,0){\makebox(0,0){\scriptsize \mbox{$#5$}}}
\end{picture}}

\def\horgraph#1#2#3#4#5#6#7{\rule[-2.8\unitlength]{0in}{5.6\unitlength}
\begin{picture}(4,4)(-#6,-#7)
\put(0,-0.1){\vector(1,0){4}}
\put(0,2){\vector(0,-1){4}}
\put(0,0.1){\vector(1,0){4}}
\put(4,2){\vector(0,-1){4}}
\put(0,-3.2){\makebox(0,0)[b]{\scriptsize \mbox{$#3$}}}
\put(4,-3.2){\makebox(0,0)[b]{\scriptsize \mbox{$#4$}}}
\put(4,2.5){\makebox(0,0)[b]{\scriptsize \mbox{$#2$}}}
\put(0,2.5){\makebox(0,0)[b]{\scriptsize \mbox{$#1$}}}
\put(2,0.5){\makebox(0,0){\scriptsize \mbox{$#5$}}}
\end{picture}}

\def\sqedges#1#2#3#4{
\begin{picture}(0,0)(0,-.3)
\put(2,-3.2){\makebox(0,0)[b]{\scriptsize \mbox{$#2$}}}
\put(4.4,0){\makebox(0,0)[l]{\scriptsize \mbox{$#4$}}}
\put(2,2.5){\makebox(0,0)[b]{\scriptsize \mbox{$#1$}}}
\put(-.4,0){\makebox(0,0)[r]{\scriptsize \mbox{$#3$}}}
\end{picture}}
\begin{document}
\title{The $A_2$ Ocneanu quantum groupo\"\i d }
\author{R. Coquereaux\\
           {\it Centre de Physique Th\'eorique - CNRS} \\
             {\it Campus de Luminy - Case 907}           \\
             {\it F-13288 Marseille - France}            }
\date{}

\maketitle
\baselineskip=11.6pt

\begin{abstract}
Description of  the smallest quantum groupo\"\i d  associated with the $A_2$ diagram. \\

\noindent
From the talk: ``Quantum groupo\"\i ds and Ocneanu bialgebras for Coxeter-Dynkin systems'' given at
the  XV Colloquio Latinoamericano de \'Algebra, Cocoyoc, M\'exico, July 20th-26th, 2003.\
{\hskip 6cm} {\tt CPT-2003/P.4514}
\end{abstract}

\baselineskip=14pt

\section{Introduction}

The Ocneanu construction associates a dialgebra  $\mathcal B$, described as a vector space with two associative $C^*$ algebra structures,  to every Dynkin diagram (of type $ADE$), more precisely to the space of paths on a given diagram. 
A general sketch of this construction was presented during several talks or conferences, since 1995,  but the first publicly available reference seems to be  \cite{Ocneanu:paths} (this hard-to-read paper,  summarising many results,  are notes taken by S. Goto,  following a set of lectures given by Prof. Ocneanu).
It is a priori clear that existence of a scalar product allows us, in each case, to trade one of the multiplications for a comultiplication, so that the obtained structure is expected to be a  bialgebra.   It was moreover expected
that  such bialgebras,  resulting from the Ocneanu construction,  were going to be weak Hopf algebras.
We believe that this property was first stated in the article  \cite{PetkovaZuber:Oc}, which
starts from conformal field theory and  provides a physical interpretation for many algebraic constructs stemming from Ocneanu's theory. 
One should notice that  the reference  \cite{Ocneanu:paths} does not mention these quantum groupo\"\i d aspects at all.

It is actually possible to derive very non-trivial results from the general construction, like the
block decomposition for the two associative structures of $\mathcal B$, 
without having to determine explicitly the matrix units  for the corresponding two products, \ie
without having to cope with the explicit definition of these two structures.
Such results  were obtained in references \cite{Coque:Qtetra}, \cite{CoqueGil:ADE}, 
\cite{CoqueGil:Tmodular} and \cite{CoqueMarina:minimal} (see also the PhD thesis \cite{Gil:thesis}), as well as in the article  \cite{PetkovaZuber:Oc} that we already quoted.
Other quantum field theoretical aspects of the construction are discussed in \cite{Pearce-Zhou} and \cite{PearceEtAl:integrable}.
However, it seems  that there is not yet  any reference providing a check that bialgebras resulting from the Ocneanu construction obey all the weak Hopf algebra axioms (which are stated, for instance, in
\cite{BohmSzlachanyi}), and I am not aware of any complete and explicit description of a single example in the available litterature.
For this reason, together with theoretical physicists colleagues from Brazil and Argentina,    we 
undertook this study and also decided to analyse explicitly, from the  weak Hopf algebra point of view, 
several examples resulting from this construction. Our joint work will appear elsewhere  \cite{CST}. 
For the purpose of these proceedings, 
I am only going to describe the smallest non-trivial example of an Ocneanu quantum groupo\"\i d, the one associated with the diagram $A_2$. 
Admitedly, this is a very simple example (maybe too simple): first of all, diagrams belonging to the $A_n$ family are of a very special kind. Moreover,  even if we restrict our attention to that family, many interesting aspects (and technical difficulties) only appear when $n>2$.  Nevertheless, even when $n=2$, the quantum groupo\"\i d  that we obtain is non-trivial (both algebra structures are isomorphic with $M(2,\CC) \oplus M(2,\CC)$) and this will allow us to see at work  many interesting features of the general theory of weak Hopf algebras, as described in \cite{BohmSzlachanyi}, \cite{NikshychVainerman} or \cite{Nill}.
The bialgebra that we discuss here is therefore only the smallest example of a very special family but the ideas  can be generalized to all Dynkin diagrams, and even to members of higher Coxeter-Dynkin systems, like the Di Francesco - Zuber graphs (which are associated with $SU(3)$ in the same way as usual Dynkin diagrams are associated with $SU(2)$), see \cite{DiF-Zub-Trieste}, \cite{DiFrancescoZuber}, \cite{Ocneanu:Bariloche}, \cite{Zuber:generaldynkin}, \cite{Zuber:Bariloche}.
 We shall not study such generalizations in the present contribution but hope that the simple example presented here will trigger the interest of the reader.

One should certainly  mention that these constructions  have a  nice interpretation \cite{Ocneanu:StFrancois} in the theory of Von Neumann algebras (study of subfactors) and can also probably be understood from the point of view of the theory of sectors, as discussed, for instance in the book \cite{EvansKawa:book}, or in references \cite{Evans}, \cite{Evans:Bariloche}, but we do not feel competent enough to discuss these aspects here. There are also interesting relations  with the theory of double categories (cf. the talk \cite{andrusk}, these proceedings).

The impatient reader may jump directly to the end of section \ref{sec:matrixrealizations} where he will find two matrices displaying matrix units for the two products.  These two expressions  therefore summarize the present paper in the most economical way.

In order to illustrate  the nice interplay between mathematics and theoretical physics, it may 
be interesting to remember that one of the motivations of the author of \cite{Ocneanu:paths} was
to recover the ADE classification of modular invariant partition functions in a particular class of two-dimensional quantum field theories \cite{cappelli}, \cite{CIZ},  \cite{Pasquier}.  

\section{Paths and essential paths}

\subsection{Diagram, vertices and edges}

We start from the Dynkin diagram $A_2$. It is displayed below. Vertices are called $\{ v_0,v_1\}$.
We also give its adjacency matrix.

\begin{figure}[hhh]
\unitlength 0.8mm
\begin{center}
\begin{picture}(55,20)(0,02)
\put(5,10){\line(1,0){15}}
\multiput(5,10)(15,0){2}{\circle*{2}}
\put(5,3){\makebox(0,0){[$v_{0}$]}}
\put(20,3){\makebox(0,0){[$v_{1}$]}}
\end{picture}
\qquad \qquad
$
{\mathcal G}_{A_2} =
\left( \begin{array}{cc}
     0 & 1 \\
     1 & 0  \\
\end{array}
\right)
$

\caption{The $A_2$ Dynkin diagram and its adjacency matrix}
\label{grA2}
\end{center}
\end{figure}
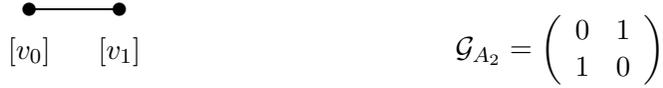

There are two oriented edges $r = v_0 \rightarrow v_1$ and $l = v_1 \rightarrow v_0$.

Elementary paths of length $0$ ($v_p  \rightarrow v_p$)  are identified with vertices  $v_p$.
Elementary paths of length $1$ are (oriented) edges.
An elementary path of length $n$ is a succession of matching edges. A path is a linear combination of elementary paths. This vector space $Paths$  comes with a prefered basis (the elementary paths) and with a scalar product (obtained by declaring that elementary paths are orthonormal). In this particularly simple case, we shall define essential paths as ``non backtracking paths''. There is a general definition of essential paths (see \cite{Ocneanu:paths}) but in the case of $A_2$ the situation is so simple that we shall use this naive definition.
Explicitly, we define essential paths as the elements of the  vector space $EssPath$ spanned by $\{v_0, v_1, r, l\}$.For a reason that will be clear later, essential paths are also called horizontal paths.
The vector space $Paths$, is graded by the length of the paths. Same thing for  its vector subspace ${\mathcal H} = EssPath =  {\mathcal H}^{0} \oplus {\mathcal H}^{1}$. 

\paragraph {Note}

In this simple example, we have a base of the space of essential paths made of vectors which are also elementary paths. This is not so for other $ADE$ diagrams, even for $A_3$!  Moreover, the expression of such basis vectors involves,  in general,  the quantum dimensions of the vertices (defined as components of the Perron Frobenius vector associated with the adjacency matrix). In the present situation, these quantum dimensions are equal to $1$.


\section{The algebra $({\mathcal B}, \circ)$}
\subsection{${\mathcal B}$ as the algebra of graded endomorphisms of ${\mathcal 
H}$}

Define ${\mathcal B} = End ({\mathcal H}^{0}) \oplus End({\mathcal H}^{1})$.
Note that $dim \,  {\mathcal B} = 2^{2} + 2^{2}=4+4=8$. It is isomorphic with $M(2,\CC) \oplus M(2,\CC)$.

\label{sec:diffusiongraphs}

We use the symbol ``hat'' to denote the dual of a vector space. The dual basis of $\{v_0, v_1, r,l\}$ is denoted  $\{\hat v_0,\hat  v_1, \hat r,\hat  l\}$. Moreover  $End H = {\mathcal H} \otimes \widehat{\mathcal H}$ and  ${\mathcal B} = 
({\mathcal H}^{0} \otimes \widehat{{\mathcal H}}^{0} )\oplus ({\mathcal H}^{1} \otimes {\widehat{\mathcal H}}^{1})
$. 
We choose the following basis in ${\mathcal B}$:
$
 \{0\hat 0,0\hat 1,1\hat 0,1\hat 1,r\hat r,r\hat  l,l\hat r,l\hat l\}
$.

Here and below we suppress a tensor product sign, so that $r \hat l$, 
for instance,  stands for $r \otimes \hat  l$.
In order to simplify the notations, we decide to call the vertices $0$ and $1$ rather than
$v_0$ and $v_1$.
It is nice to display these elements $u\hat v \doteq u \otimes \hat v$ as 
vertical diffusion graphs.
$$
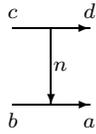
\setlength{\unitlength}{.1in}
\vertgraph {c}{d}{b}{a}{ n}{0}{.5}\punit2
$$
Rather than using diffusion graphs, as above, one could use dual pictures and replace triple points (vertices) by triangles. A diffusion graphs then becomes a set of two triangles glued along a common basis. For this reason, the object $\mathcal B$ was nicknamed DTA in reference \cite{Ocneanu:paths}, the acronym standing for ``Double Triangle Algebra''.

\subsection{Identification of ${\widehat{\mathcal H}}$ and ${\mathcal H}$}

Since $\mathcal H$ is endowed with a particular  scalar product, we 
can identify  it with its dual. Hence we can aslo
identify $End H = {\mathcal H} \otimes \widehat{\mathcal H}$ with 
${\mathcal H} \otimes {\mathcal H}$. In particular we identify ${\mathcal B} $ 
with 
$$
({\mathcal H}^{0} \otimes {\mathcal H}^{0} )\oplus ({\mathcal H}^{1} \otimes {\mathcal H}^{1})
$$
For this reason, we simplify the notation given for the chosen base of $\mathcal B$ and write simply (drop the hats): 
$
\{0 0,0 1,1 0,1 1,r r,r  l,l r,l l\}
$.

Warning: Although we identify $\mathcal H$ with ${\widehat{\mathcal H}}$,
 we do \underline{not} identify the algebra $\mathcal B$ with its dual ${\widehat{\mathcal B}}$ (at least, not yet).
\paragraph{Note} For more general diagrams, the same kind of identification holds as well, and comes
from the fact that the vector space $Paths$ always comes with a preferred basis and therefore with a canonical scalar product. However it is almost always dangerous (besides  the case of $A_2$!) to identify  $\mathcal B$ and ${\widehat{\mathcal B}}$.

\subsection{Diffusion graphs}

Diffusion graphs of length $n=0$ 
$$
0  0
\quad
\equiv 
\setlength{\unitlength}{.1in}
\vertgraph {0}{0}{0}{0}{ 0}{0}{.5}\punit2, \,
0 1
\quad
\equiv 
\setlength{\unitlength}{.1in}
\vertgraph {0}{0}{1}{1}{ 0}{0}{.5}\punit2,  \,
1 0 
\quad
\equiv 
\setlength{\unitlength}{.1in}
\vertgraph {1}{1}{0}{0}{ 0}{0}{.5}\punit2,  \,
1 1
\quad
 \equiv 
\setlength{\unitlength}{.1in}
\vertgraph {1}{1}{1}{1}{ 0}{0}{.5}\punit2, 
$$
Diffusion graphs of length $n=1$:
$$
r r
\quad
\equiv 
\setlength{\unitlength}{.1in}
\vertgraph {0}{1}{0}{1}{ 1}{0}{.5}\punit2, \,
r l
\quad
\equiv 
\setlength{\unitlength}{.1in}
\vertgraph {0}{1}{1}{0}{ 1}{0}{.5}\punit2,  \,
l r
\quad
\equiv 
\setlength{\unitlength}{.1in}
\vertgraph {1}{0}{0}{1}{ 1}{0}{.5}\punit2,  \,
l l
\quad
 \equiv 
\setlength{\unitlength}{.1in}
\vertgraph {1}{0}{1}{0}{ 1}{0}{.5}\punit2, 
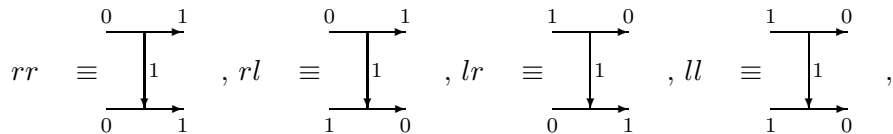
$$

\subsection{The (first) product on  ${\mathcal B}$ : composition of 
endomorphisms}

Let $u \otimes v$ and  $u' \otimes v'$ in ${\mathcal B}$. We define
$$
(u \otimes v)\circ (u' \otimes v') = \left< v,u' \right>  u \otimes v'
$$

Here the complex number $\left<  v,u' \right>$ denotes the pairing between $ {\mathcal H}$ and its dual. 
The non-zero pairings\footnote{They can be interpreted as non-zero scalar products $(v,u')$ between orthonormal basis elements in $\mathcal H$.} between basis elements are
$$
\left<  0,0 \right> = \left< 1,1 \right> =\left< l,l \right> =\left< 
r,r \right> = 1
$$
The non-zero products are therefore given by the following table:
 $$
    \begin{array}{c|cccccccc}
    \circ \nearrow & 00 & 01 & 10 & 11 & rr & rl & lr & ll \\
     \hline \\
    00  & 00 & 01 & . & . & . & . & . & . \\
    01  & . & . & 00 & 01 & . & . & . & . \\
    10  & 10 & 11 & . & . & . & . & . & . \\
    11  & . & . & 10 & 11 & . & . & . & . \\ 
    rr  & . & . & . & . & rr & rl & . & . \\
    rl  & . & . & . & . & . & . & rr & rl \\ 
    lr  & . & . & . & . & lr & ll & . & . \\
    ll  & . & . & . & . & . & . & lr & ll \\
    \end{array}
$$

\paragraph{The unit.}
The neutral element for the above product  is
$$
\one = 00 + 11 + rr + ll
$$

\subsection{ $({\mathcal B}, \circ)$ as a sum of blocks}

The next discussion is therefore  rather trivial since we already know from the very definition of ${\mathcal B} $ that it is isomorphic with $M(2,\CC) \oplus M(2,\CC)$  but several ingredients of this discussion will be useful later.

\paragraph{Minimal central projectors for $({\mathcal B}, \circ)$}

From the multiplication table, we can check that 
\begin{eqnarray*}
\pi_{0} &=& 00 + 11 \quad \mbox{and} \\
\pi_{1} &=& rr + ll
 \end{eqnarray*}
  are minimal  central projectors. Moreover they are orthogonal and their sum is 
$\one$. We define the ideals $\pi_{0} {\mathcal B} = \pi_{0} {\mathcal B} \pi_{0}$ and
 $\pi_{1} {\mathcal B} = \pi_{1} {\mathcal B} \pi_{1}$. These will be the two blocks.
\begin{eqnarray*}
 {\mathcal B}_{0} &=& \pi_{0} {\mathcal B} \quad  \mbox{is the linear span of}\; \{00,01,10,11\} \\
 {\mathcal B}_{1} &=& \pi_{1} {\mathcal B} \quad  \mbox{is the linear span of} \; \{rr,rl,lr,ll\}
 \end{eqnarray*}
 As expected these two blocks are $4$-dimensional, 
the algebra ${\mathcal B}$ is semi-simple and  we can write each of the two blocks as $2\times 
 2$ complex matrices.
 
 \paragraph{Matrix units for $({\mathcal B}, \circ)$}
 We identify the elements of the basis $\{00,01,10,11; rr,rl,lr,ll\}$ of ${\mathcal B}$  with matrix units for the blocks
 ${\mathcal B}_0$ and ${\mathcal B}_1$ as follows:
 
 {\tiny
 $
 \begin{array}{cc}
\begin{array}{cc}
   
00 =\left(
\begin{array}{c|c}
    \begin{array}{cc}
        1 & 0  \\
        0 & 0
    \end{array}
     & 0  \\ \hline \\
    0 & 
    \begin{array}{cc}
        0 & 0  \\
        0 & 0
    \end{array}
\end{array}
\right)

&

01 = \left(
\begin{array}{c|c}
    \begin{array}{cc}
        0 & 1  \\
        0 & 0
    \end{array}
     & 0  \\ \hline \\
    0 & 
    \begin{array}{cc}
        0 & 0  \\
        0 & 0
    \end{array}
\end{array}
\right)

\\

&

\\

10 = \left(
\begin{array}{c|c}
    \begin{array}{cc}
        0 & 0  \\
        1 & 0
    \end{array}
     & 0  \\ \hline \\
    0 & 
    \begin{array}{cc}
        0 & 0  \\
        0 & 0
    \end{array}
\end{array}
\right)

&

11 = \left(
\begin{array}{c|c}
    \begin{array}{cc}
        0 & 0  \\
        0 & 1
    \end{array}
     & 0  \\  \hline \\
    0 & 
    \begin{array}{cc}
        0 & 0  \\
        0 & 0
    \end{array}
\end{array}
\right)

\end{array}

&

\begin{array}{cc}
    
rr = \left(
\begin{array}{c|c}
    \begin{array}{cc}
        0 & 0  \\
        0 & 0
    \end{array}
     & 0  \\  \hline \\
    0 & 
    \begin{array}{cc}
        1 & 0  \\
        0 & 0
    \end{array}
\end{array}
 \right)

&

rl=\left(
\begin{array}{c|c}
    \begin{array}{cc}
        0 & 0  \\
        0 & 0
    \end{array}
     & 0  \\  \hline \\
    0 & 
    \begin{array}{cc}
        0 & 1  \\
        0 & 0
    \end{array}
\end{array}
\right)

\\

lr=\left(
\begin{array}{c|c}
    \begin{array}{cc}
        0 & 0  \\
        0 & 0
    \end{array}
     & 0  \\  \hline \\
    0 & 
    \begin{array}{cc}
        0 & 0  \\
        1 & 0
    \end{array}
\end{array}
\right)

&

ll=\left(
\begin{array}{c|c}
    \begin{array}{cc}
        0 & 0  \\
        0 & 0
    \end{array}
     & 0  \\  \hline \\
    0 & 
    \begin{array}{cc}
        0 & 0  \\
        0 & 1
    \end{array}
\end{array}
\right)
\end{array}
\end{array}
$
}

 Matrix units of a block ${\mathcal B}_\alpha$  span the one dimensional subspace  $q_{i}^{\alpha} {\mathcal B}_{\alpha} q_{j}^{\alpha}$ where  the two $q$'s are minimal (but non central) projectors. For instance,
 $$
 {\mathcal B}_{\alpha} = q_{1}^{\alpha} {\mathcal B}_{0} q_{1}^{\alpha} +  q_{1}^{\alpha} {\mathcal B}_{0} q_{2}^{\alpha} +  q_{2}^{\alpha} {\mathcal B}_{0}  q_{1}^{\alpha} +  q_{2}^{\alpha} {\mathcal B}_{0} q_{2}^{\alpha}
 $$
where we take $q_{1}^{0} = 00$, $q_{2}^{0} = 11$, so that 
\begin{eqnarray*}
00 \circ \{00,01,10,11\} \circ 00  = \{00 \}  &{}&
00 \circ \{00,01,10,11\} \circ 11 = \{01 \}  \\
11  \circ \{00,01,10,11\} \circ 00 = \{10 \}   &{}&
 11   \circ \{00,01,10,11\} \circ 11 = \{11 \} \\
\end{eqnarray*}
The already defined central minimal projector relative to the block ${\mathcal B}_{\alpha}$ is recovered as 
 $\pi_{\alpha} = q_{1}^{\alpha} \oplus q_{2}^{\alpha}$.

  Of course, such results were a priori obvious since the 
algebra elements have been defined as endomorphisms of a graded vector 
space. The point that we want to make is that matrix units relative to a given block (a given length) and 
associated with bilateral ideals of $({\mathcal B}, \circ)$ are naturally associated with pairs of essential paths (specifying one line and one column). The whole multiplicative structure can be encoded by the following picture where each entry denotes a single matrix unit (set the others entries to zero):

$
\begin{array}{ccc}

\left(
    \begin{array}{cc}
     00 & 01 \\
        {}&{} \\
     10 & 11
    \end{array}
\right)
\oplus
\left(
     \begin{array}{cc}
    rr & rl \\
        {}&{} \\
     lr & ll
    \end{array}
\right)

& \equiv &

\left(
    \begin{array}{cc}
        \setlength{\unitlength}{.1in}
\vertgraph {0}{0}{0}{0}{ 0}{0}{.5}\punit2 &  \setlength{\unitlength}{.1in}\vertgraph {0}{0}{1}{1}{ 0}{0}{.5}\punit2 \\
        {}&{} \\
 \setlength{\unitlength}{.1in}\vertgraph {1}{1}{0}{0}{ 0}{0}{.5}\punit2 &  \setlength{\unitlength}{.1in} \vertgraph {1}{1}{1}{1}{ 0}{0}{.5}\punit2
    \end{array}
\right)
\oplus
\left(
    \begin{array}{cc}
        \setlength{\unitlength}{.1in}
\vertgraph {0}{1}{0}{1}{ 1}{0}{.5}\punit2 &  \setlength{\unitlength}{.1in}\vertgraph {0}{1}{1}{0}{ 1}{0}{.5}\punit2 \\
        {}&{} \\
 \setlength{\unitlength}{.1in}\vertgraph {1}{0}{0}{1}{ 1}{0}{.5}\punit2 &  \setlength{\unitlength}{.1in} \vertgraph {1}{0}{1}{0}{ 1}{0}{.5}\punit2
    \end{array}
\right)

\end{array}
$
The multiplication $\circ$ for the product of matrix units becomes intuitive when we compose vertically these diffusion diagrams.

\subsection{The cogebra  $({\mathcal B}, \Delta)$}

We now define a coproduct on the vector space ${\mathcal B}$. It will be compatible with the already defined  product $\circ$. Therefore ${\mathcal B}$ turns out to be a bialgebra.

\subsubsection{Definition}

 The expression of the product $\circ$ comes directly from the definition of $\mathcal B$ as a (graded) algebra of endomorphisms. The situation is not that simple for the coproduct $\Delta$: there are several ways to obtain the following definition from general grounds but we postpone this discussion to the end of section \ref{sec:cells} (see also the end of section \ref{sec:concat}).
  At the moment we define explicitly this non-trivial coproduct $\Delta: {\mathcal B} \mapsto  {\mathcal B} \otimes {\mathcal B}$ by its action on the chosen basis elements of $\mathcal B$ and extend it by linearity.

\begin{center}
 $
 \begin{array}{cc}
\begin{array}{ccc}
\Delta 00 &=& 00 \bigotimes 00   + rr \bigotimes  ll \\
\Delta 01 &=& 01 \bigotimes 01   + rl \bigotimes  lr \\
\Delta 10 &=& 10 \bigotimes 10   + lr \bigotimes  rl \\
\Delta 11 &=& 11 \bigotimes 11   + ll \bigotimes  rr \\
\end{array}
& \qquad
\begin{array}{ccc}
\Delta rr &=& 00 \bigotimes rr   + rr \bigotimes  11 \\
\Delta rl &=& 01 \bigotimes rl   + rl \bigotimes  10 \\
\Delta lr &=& 10 \bigotimes lr   + lr \bigotimes  01 \\
\Delta ll &=& 11 \bigotimes ll   + ll \bigotimes  00 \\
\end{array}
\end{array}
$
\end{center}

\subsubsection{Compatibility with the algebra structure}
This  coproduct is compatible with the already defined multiplication $\circ$ on $ {\mathcal B}$ : it is  an homomorphism from  this algebra to its tensor square (the natural multiplication in the later being
$(a\otimes b)\circ (c \otimes d) = (a \circ c) \otimes (b \circ d)$). The compatibility property  can be  explicitly
checked and reads:
$\Delta (a \circ b) = \Delta a \circ \Delta b$.
Conclusion: $(\mathcal B, \circ, \Delta)$ is a bialgebra.

\subsubsection{The coproduct of the unit}
 \label{sec:deltaone}
From the above definition we  calculate the expression of $\Delta \one$, where $\one = 00 + 11 + rr + ll$ and find
$$\Delta \one = (00 + ll) \otimes (00 + rr) + (rr + 11) \otimes (ll + 
11)$$
From the fact that  $\Delta \one$ is not equal to $\one \otimes \one$ we see that  $\mathcal B$ cannot be a Hopf algebra. However, we shall see later that it is a weak Hopf algebra (a quantum groupo\"\i d ).

\subsubsection{Counit}
\label{sec:counit}
The couint $\epsilon$, for a bialgebra, should be such that the usual commutative diagrams hold. In the present case, we  have a system of $4+4$ equations to solve.  The first four are
\begin{center}
$
\begin{array}{cc}
\begin{array}{ccc}
\epsilon(00) 00 + \epsilon(rr) ll &=& 00 \\
\epsilon(10) 10 + \epsilon(lr) rl &=& 10 \\
\end{array}
&
\begin{array}{ccc}
\epsilon(01) 01 + \epsilon(rl) lr &=& 01 \\
\epsilon(11) 11 + \epsilon(ll) rr &=& 11 
\end{array}
\end{array}
$
 \end{center}
 
whose solution is\footnote{One can check that the other equations are satisfied as well.}
\begin{eqnarray*}
\epsilon(00)=\epsilon(01)=\epsilon(10)=\epsilon(11)=1\\
\epsilon(rr)=\epsilon(rl)=\epsilon(lr)=\epsilon(ll)=0
\end{eqnarray*}
Therefore $(\mathcal B, \circ, \Delta, \one, \epsilon)$ is a unital and counital bialgebra.

 For non-trivial  quantum groupo\"\i ds, the counit is not a homomorphism. Another property (replacing the homomorphims property)  has to be satisfied. We return later to this point.

\section{The dual algebra $(\widehat {\mathcal B}, \, {\widehat \circ} \,)$}

We could very well carry out the whole study at the level of the space $ {\mathcal B}$, endowed with  a product and a coproduct, but we prefer -- maybe a matter of taste -- to replace the coproduct  on  $ {\mathcal B}$ by a product on its dual.
Using the same basis as before on  $ {\mathcal B}$, we denote  the   dual basis as follows:
 $\{\widehat{00}, \widehat{01}, \widehat{10}, \widehat{11}, \widehat{rr}, \widehat{rl}, \widehat{lr}, \widehat{ll}\}$

\subsection{The product  $\, {\widehat \circ} \,$ on $\widehat {\mathcal B}$ }

Since we have a coproduct $\Delta$ on the vector space $ {\mathcal B}$, we have a product on its dual $\widehat {\mathcal B}$.  Remark: this correspondance is canonical and does not involve any choice of scalar product.  
From the expression of $\Delta$, we obtain the following multiplication table, written in terms of the dual basis defined before:
  
$$
\begin{array}{c|cccccccc}
    \, {\widehat \circ} \, \nearrow & \widehat{00} & \widehat{01} & \widehat{10} & \widehat{11} & \widehat{rr} & \widehat{rl} & \widehat{lr} & \widehat{ll} \\
    \hline \\
    \widehat{00} & \widehat{00} & . & . & . & \widehat{rr} & . & . & . \\
    \widehat{01} & . & \widehat{01} & . & . & . & \widehat{rl} & . & . \\ 
     \widehat{10} & . & . & \widehat{10} & . & . & . & \widehat{lr} & . \\ 
      \widehat{11} & . & . & . & \widehat{11} & . & . & . & \widehat{ll} \\ 
       \widehat{rr} & . & . & . & \widehat{rr} & . & . & . & \widehat{00} \\ 
        \widehat{rl} & . & . & \widehat{rl} & . & . & . & \widehat{01} & . \\ 
	 \widehat{lr} & . & \widehat{lr} & . & . & . & \widehat{10} & . & . \\ 
	  \widehat{ll} & \widehat{ll} & . & . & . & \widehat{11} & . &  & . 
    \end{array}
  $$
  
\paragraph{The unit of $\, {\widehat \circ} \,$}
From the above table, we read the expression of the neutral element  $\widehat \one$ 
$$
\widehat \one = \widehat{00} + \widehat{01} + \widehat{10} + \widehat{11}
$$
 This was already clear (or conversely)  from the expression of the counit $\epsilon$ given in section \ref{sec:counit}

 \subsection{$(\widehat {\mathcal B}, \, {\widehat \circ} \,)$ as a sum of blocks}

\paragraph{Minimal central projectors and blocks.}
From the multiplication table, we see that
$$\varpi_{0} = \widehat{00} + \widehat{11} \quad \mbox{and}
 \quad \varpi_{1} = \widehat{01} + \widehat{10}$$
are minimal central projectors. Moreover they are orthogonal and their sum is $\widehat \one$.
The two blocks (of dimension 4) are the ideals $\varpi_{0} {\widehat {\mathcal B}}  = \{ \widehat{00}, \widehat{11}, \widehat{rr}, \widehat{ll} \}$ and $ \varpi_{1} {\widehat {\mathcal B}}  =  \{ \widehat{01}, \widehat{10}, \widehat{rl}, \widehat{lr}\}$.

\paragraph{Matrix units.}

We now  find  (non central) minimal projectors and decompose each 
 of the two blocks of this semi-simple algebra on the corresponding matrix units.
Call ${\widehat {\mathcal B}}_{ 0} = \varpi_{0} {\widehat {\mathcal B}}$. Notice that $\widehat{00}$ and $\widehat{11}$ are projectors, but $\widehat{rr}$ and $\widehat{ll}$ are not.
These two projectors are supplementary: 
 ${\widehat {\mathcal B}}_{ 0} =  {\widehat {\mathcal B}}_{ 0} \, {\widehat \circ} \, \widehat{00} \oplus  {\widehat {\mathcal B}}_{ 0} \, {\widehat \circ} \, \widehat{11} = \{\widehat{00},\widehat{ll}\} 
 \oplus \{\widehat{11},\widehat{rr}\}.$
 When expressed in terms of matrices; ${\widehat {\mathcal B}}_{ 0}$ will be the first 
 block (a $2\times 2$ matrix), 
 ${\widehat {\mathcal B}}_{ 0} \, {\widehat \circ} \, \widehat{00}$ will be the first column of this block and ${\widehat {\mathcal B}}_{ 0} 
 \, {\widehat \circ} \, \widehat{11}$ the second column.
 More precisely, we have the following matrix structure for ${\widehat {\mathcal B}}_{ 
 0}  $

 $$
\left(
 \begin{array}{c|c }
 \widehat{00} \, {\widehat \circ} \, {\widehat {\mathcal B}}_{ 0}\, {\widehat \circ} \, \widehat{00} = \{\widehat{00}\} &  \widehat{00} \, {\widehat \circ} \, {\widehat {\mathcal B}}_{ 0}  
 \, {\widehat \circ} \, \widehat{11} = \{\widehat{rr}\} \\ \hline
 \widehat{11} \, {\widehat \circ} \, {\widehat {\mathcal B}}_{ 0}\, {\widehat \circ} \, \widehat{00} = \{\widehat{ll}\} &  \widehat{11}\, {\widehat \circ} \, {\widehat {\mathcal B}}_{ 0} \, {\widehat \circ} \, \widehat{11} = \{\widehat{11}\} \\
  \end{array}
  \right)
$$

Call ${\widehat {\mathcal B}}_{ 1} = \varpi_{1} {\widehat {\mathcal B}}$.
Notice that $\widehat{01}$ and $\widehat{10}$ are projectors, but $\widehat{lr}$ and $\widehat{rl}$ are not.
These two projectors are supplementary: 
 ${\widehat {\mathcal B}}_{ 1} =  {\widehat {\mathcal B}}_{ 1} \, {\widehat \circ} \, \widehat{01} \oplus  {\widehat {\mathcal B}}_{ 1}\, {\widehat \circ} \, \widehat{10} = \{\widehat{01},\widehat{lr}\} 
 \oplus \{\widehat{10},\widehat{rl}\}$
 When expressed in terms of matrices; ${\widehat {\mathcal B}}_{ 0}$ will be the second
 block (a $2\times 2$ matrix), 
 ${\widehat {\mathcal B}}_{ 1}\, {\widehat \circ} \, \widehat{01}$ will be the first column of this block (${\widehat {\mathcal B}}_{ 1} 
 \, {\widehat \circ} \, \widehat{10}$ will be the second column).
 More precisely, we have the following matrix structure for ${\widehat {\mathcal B}}_{ 
 1}  $
 $$
 \left(
 \begin{array}{c | c }
 \widehat{01}\, {\widehat \circ} \, {\widehat {\mathcal B}}_{ 1}\, {\widehat \circ} \, \widehat{01} = \{\widehat{01}\} &  \widehat{01}\, {\widehat \circ} \, {\widehat {\mathcal B}}_{ 1}\, {\widehat \circ} \, \widehat{10} = \{\widehat{rl}\} \\
\hline
 \widehat{10}\, {\widehat \circ} \, {\widehat {\mathcal B}}_{ 1}\, {\widehat \circ} \, \widehat{01} = \{\widehat{lr}\} &  \widehat{10}\, {\widehat \circ} \, {\widehat {\mathcal B}}_{ 1}\, {\widehat \circ} \, \widehat{10} = 
 \{\widehat{10}\} \\  
   \end{array}
     \right)
 $$

  We therefore  identify the elements of the basis $\{\widehat{00}, \widehat{01}, \widehat{10}, \widehat{11}, \widehat{rr}, \widehat{rl}, \widehat{lr}, \widehat{ll}\}$ of ${\widehat {\mathcal B}}$  with matrix units for the blocks
 ${\widehat {\mathcal B}}_{ 0}$ and ${\widehat {\mathcal B}}_{ 1}$ as follows:
 
 {\tiny
 $
 \begin{array}{cc}
\begin{array}{cc}
   
\widehat{00} =\left(
\begin{array}{c|c}
    \begin{array}{cc}
        1 & 0  \\
        0 & 0
    \end{array}
     & 0  \\ \hline \\
    0 & 
    \begin{array}{cc}
        0 & 0  \\
        0 & 0
    \end{array}
\end{array}
\right)

&

\widehat{rr} = \left(
\begin{array}{c|c}
    \begin{array}{cc}
        0 & 1  \\
        0 & 0
    \end{array}
     & 0  \\ \hline \\
    0 & 
    \begin{array}{cc}
        0 & 0  \\
        0 & 0
    \end{array}
\end{array}
\right)

\\

&

\\

\widehat{ll} = \left(
\begin{array}{c|c}
    \begin{array}{cc}
        0 & 0  \\
        1 & 0
    \end{array}
     & 0  \\ \hline \\
    0 & 
    \begin{array}{cc}
        0 & 0  \\
        0 & 0
    \end{array}
\end{array}
\right)

&

\widehat{11} = \left(
\begin{array}{c|c}
    \begin{array}{cc}
        0 & 0  \\
        0 & 1
    \end{array}
     & 0  \\  \hline \\
    0 & 
    \begin{array}{cc}
        0 & 0  \\
        0 & 0
    \end{array}
\end{array}
\right)

\end{array}

&

\begin{array}{cc}
    
\widehat{01} = \left(
\begin{array}{c|c}
    \begin{array}{cc}
        0 & 0  \\
        0 & 0
    \end{array}
     & 0  \\  \hline \\
    0 & 
    \begin{array}{cc}
        1 & 0  \\
        0 & 0
    \end{array}
\end{array}
 \right)

&

\widehat{rl}=\left(
\begin{array}{c|c}
    \begin{array}{cc}
        0 & 0  \\
        0 & 0
    \end{array}
     & 0  \\  \hline \\
    0 & 
    \begin{array}{cc}
        0 & 1  \\
        0 & 0
    \end{array}
\end{array}
\right)

\\

\widehat{lr}=\left(
\begin{array}{c|c}
    \begin{array}{cc}
        0 & 0  \\
        0 & 0
    \end{array}
     & 0  \\  \hline \\
    0 & 
    \begin{array}{cc}
        0 & 0  \\
        1 & 0
    \end{array}
\end{array}
\right)

&

\widehat{10}=\left(
\begin{array}{c|c}
    \begin{array}{cc}
        0 & 0  \\
        0 & 0
    \end{array}
     & 0  \\  \hline \\
    0 & 
    \begin{array}{cc}
        0 & 0  \\
        0 & 1
    \end{array}
\end{array}
\right)
\end{array}
\end{array}
$
}

\paragraph{Note.} This calculation shows that, for  $A_2$, the dual of the basis of matrix units for the multiplication $\circ$ on $\mathcal{B}$ is also a basis of matrix units (but not in the same order) for the multiplication $\, {\widehat \circ} \,$ on $\widehat{\mathcal{B}}$. This is however a particular feature of the $A_2$ case (it is not even valid for $A_3$). Later we shall interpret this property by saying that non-trivial ``cells'',  in the present case, have a value equal to $1$.

\subsection{Vertical space and dual diffusion graphs}

In the case of $\mathcal B$, we started from a vector space $\mathcal H$, called ``horizontal'' , whose elements
were interpreted as essential paths on the given diagram, and $\mathcal B$ was defined as the algebra $End \, {\mathcal H}^0 \oplus End \,  {\mathcal H}^1$.
In the case of $\widehat{\mathcal B}$, the situation is similar but the starting point is different: the multiplication table is somehow given and the algebra is  the sum of two simple blocks; it can therefore be written as $End \,  {\mathcal V}^0 \oplus End \,  {\mathcal V}^1$. The vector space ${\mathcal V} = {\mathcal V}^0 \oplus {\mathcal V}^1$ is called vertical and its elements are called ``vertical paths''.
Again, the matrix units should be naturally associated with pairs of basis vectors
specifying the corresponding line and column of the given block  ($x=0$ or $1$). 
In the case of  $\circ$, blocks were labelled with an integer $n =0$ or $1$ called the ``horizontal length''. In the case of $\, {\widehat \circ} \,$, by analogy, the label $x$ can be called ``vertical length''.
It is convenient  to display the corresponding matrix units  as 
horizontal diffusion graphs:

$$
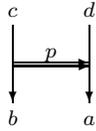
\setlength{\unitlength}{.1in}
\horgraph {c}{d}{b}{a}{ p}{0}{.5}\punit2
$$

Dual diffusion graphs ``of length'' $x=0$ (first block):

$$
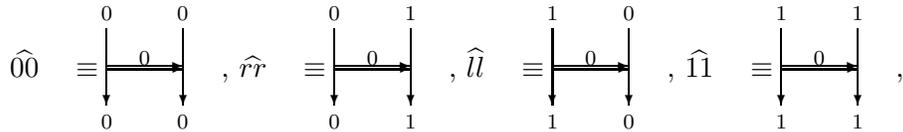
\widehat{00}
\quad
\equiv 
\setlength{\unitlength}{.1in}
\horgraph {0}{0}{0}{0}{ 0}{0}{.5}\punit2, \,
\widehat{rr}
\quad
\equiv 
\setlength{\unitlength}{.1in}
\horgraph {0}{1}{0}{1}{ 0}{0}{.5}\punit2,  \,
\widehat{ll}
\quad
\equiv 
\setlength{\unitlength}{.1in}
\horgraph {1}{0}{1}{0}{ 0}{0}{.5}\punit2,  \,
\widehat{11}
\quad
 \equiv 
\setlength{\unitlength}{.1in}
\horgraph {1}{1}{1}{1}{ 0}{0}{.5}\punit2, 
$$

Dual diffusion graphs ``of length'' $x=1$ (second  block):

$$
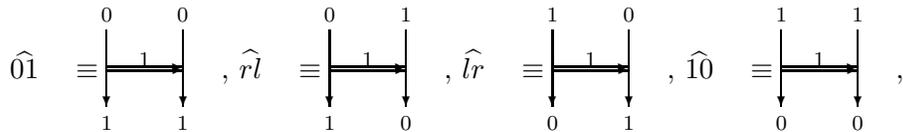
\widehat{01}
\quad
\equiv 
\setlength{\unitlength}{.1in}
\horgraph {0}{0}{1}{1}{ 1}{0}{.5}\punit2, \,
\widehat{rl}
\quad
\equiv 
\setlength{\unitlength}{.1in}
\horgraph {0}{1}{1}{0}{ 1}{0}{.5}\punit2,  \,
\widehat{lr}
\quad
\equiv 
\setlength{\unitlength}{.1in}
\horgraph {1}{0}{0}{1}{ 1}{0}{.5}\punit2,  \,
\widehat{10}
\quad
 \equiv 
\setlength{\unitlength}{.1in}
\horgraph {1}{1}{0}{0}{ 1}{0}{.5}\punit2, 
$$

The reader will have noticed that these graphs have exactely the same endpoints as those
given in section \ref{sec:diffusiongraphs},  they  only differ by their edges. The internal ones  are now drawn from left to right rather than from top to bottom.

Remark : at the moment,  we have not identified  the vector space $\mathcal B$ with its dual, so that there is no way to relate linearly the vertical diffusion graphs (matrix units for the multiplication $\circ$) with the horizontal diffusion graphs  (matrix units for the multiplication $\, {\widehat \circ} \,$).
Notice that we decide, in general, to label the later with linear combinations of elements belonging to the dual basis of the former;  so, in a sense, we are not using the  ``vertical paths'',  but we could have made a different choice: our labels have an intuitive meaning when read from top to bottom, but not from left to right.

\paragraph{Matrix units (again).}
The whole multiplicative structure can be encoded by the following picture where each entry denotes a single matrix unit (set the others entries to zero):

$
\begin{array}{ccc}

\left(
    \begin{array}{cc}
     {\widehat {00}} &  {\widehat {rr}} \\
        {}&{} \\
      {\widehat  {ll}} &  {\widehat  {11}}
    \end{array}
\right)
\oplus
\left(
     \begin{array}{cc}
     {\widehat {01}} &  {\widehat { rl}} \\
        {}&{} \\
      {\widehat { lr}} &  {\widehat {10}}
    \end{array}
\right)

& \equiv &

\left(
    \begin{array}{cc}
        \setlength{\unitlength}{.1in}
\horgraph {0}{0}{0}{0}{ 0}{0}{.5}\punit2 &  \setlength{\unitlength}{.1in}\horgraph {0}{1}{0}{1}{ 0}{0}{.5}\punit2 \\
        {}&{} \\
 \setlength{\unitlength}{.1in}\horgraph {1}{0}{1}{0}{ 0}{0}{.5}\punit2 &  \setlength{\unitlength}{.1in} \horgraph {1}{1}{1}{1}{ 0}{0}{.5}\punit2
    \end{array}
\right)
\oplus
\left(
    \begin{array}{cc}
        \setlength{\unitlength}{.1in}
\horgraph {0}{0}{1}{1}{ 1}{0}{.5}\punit2 &  \setlength{\unitlength}{.1in}\horgraph {0}{1}{1}{0}{ 1}{0}{.5}\punit2 \\
        {}&{} \\
 \setlength{\unitlength}{.1in}\horgraph {1}{0}{0}{1}{ 1}{0}{.5}\punit2 &  \setlength{\unitlength}{.1in} \horgraph {1}{1}{0}{0}{ 1}{0}{.5}\punit2
    \end{array}
\right)

\end{array}
$

The multiplication $\, {\widehat \circ} \,$ for the product of matrix units is intuitive when we compose horizontally these diagrams.

\section{Cells} 
\label{sec:cells}
The original description of Ocneanu cells (and cell systems) was developped in the context of the theory of  paragroups (see reference \cite{Ocneanu:paragroups}). It  is not the right place, here, to give a detailed account of this theory but we shall nevertheless freely use the corresponding terminology that we adapt for our purpose. 
In general a cell is characterized by its top and bottom labels, which are horizontal paths ($\xi$, $\eta$) corresponding
to a block $n$ of $\mathcal B$ (essential paths of  length $n$), and by its  left and right labels, corresponding to a block $x$ of $\widehat{\mathcal B}$ (vertical paths $\alpha$, $\beta$ ``of length $x$''). Graphically, we are pairing a vertical diffusion diagram with an horizontal diffusion diagram and the result depends therefore on the choice of two horizontal paths and two vertical paths (hence the name ``cells'').  Formally, there are two kinds of cells: those obtained by pairing the matrix units of the $\, {\widehat \circ} \,$ product (in $\widehat{\mathcal B}$) with the matrix units of the $\circ$ product (in $\mathcal B$), and those obtained by pairing the two corresponding dual basis. In the case of $A_2$, this distinction is not important (the numerical values\footnote{Sometimes the word ``cell''  is used to denote the above complex numbers but sometimes the parameter  $x$ is kept free,  and  the above  value is called the ``value of the connection $x$'' on the given cell.} are the same anyway).  Cells obey a number of axioms or properties  (see \cite{Ocneanu:paragroups},  \cite{Ocneanu:paths} or \cite{CST}) for details); one important property
is that cells can be composed (horizontally or vertically) like the Boltzman weights of statistical mechanics. In other words, one can calculate all the cells from the values of  ``basic cells''  (which have $n=1$ and $x=1$).  
Cells are displayed as follows (the $v_i$'s denote the source and target of the corresponding paths)
$$
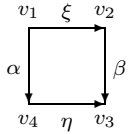
\setlength{\unitlength}{.1in}
\sqedges {\xi }{\eta}{\alpha}{\beta }
\sqface {v_{4}}{v_{3}}{v_{2}}{v_{1}}{}{0}{.5}\punit2
$$
Notice that if we graphically replace the diffusion graphs by the dual representation using double triangles, we see that the pairing of two double triangles is represented by a tetrahedron (with a special coloring of vertices);  
this shows that  Ocneanu cells can be considered as a generalizations of $6j$ symbols.

If the structure constants of the coproduct $\Delta$ in $\mathcal B$  (or of the product $\, {\widehat \circ} \,$ in $\widehat{\mathcal B}$) are somehow given -- and this was our attitude in this article -- , the values of cells can be calculated. In particular, in the case of $A_2$, there are only two basic cells (which are mirror images): from the fact that 
$ \widehat{rl}$ and $ \widehat{lr}$ are matrix units for $\, {\widehat \circ} \,$, and the fact that 
$
< \widehat{rl}, rl > = 1, 
< \widehat{lr}, lr > = 1
$
we see that 
$$
\setlength{\unitlength}{.1in}
\sqedges {r}{l}{r}{l }
\sqface {1}{0}{1}{0}{}{0}{.5}\punit2
\quad = \qquad
\setlength{\unitlength}{.1in}
\sqedges {l }{r}{l}{r }
\sqface {0}{1}{0}{1}{}{0}{.5}\punit2
\quad = \qquad 1
$$

However, in practice, we follow\footnote{See also the end of section \ref{sec:concat}} an opposite approach. Indeed, the difficulty, for general diagrams, is to obtain the definition of $\Delta$ (or of $\, {\widehat \circ} \,$) and the only general method that we know how to generalize is to start from the basic cells (for which a general formula can be produced), to calculate all the other cells by using horizontal or vertical composition of cells, and to use the result of this calculation to deduce the expression of  matrix units for $\, {\widehat \circ} \,$, in terms of the (dual of the)  matrix units for $\circ$.

\section{From bialgebras to dialgebras}

 We have a product $\circ$, a compatible coproduct $\Delta$, a unit $\one$, a counit $\epsilon$ and an antipode $S$ (as we shall see later) on the space $\mathcal B$.
By duality, we have a coproduct ${\widehat \Delta}$, a compatible  product $\, {\widehat \circ} \,$,  a counit $\widehat \epsilon$,
a unit $\widehat \one$ and an antipode $\widehat S$ on its dual  $\widehat {\mathcal B}$.

The {\sl choice\/} of a non degenerated scalar product in $\mathcal B$ allows one to identify this vector space with its dual and, as a consequence, to define for instance both multiplications on the same space. This procedure is actually rather tricky in general (\ie for diagrams more complicated than $A_2$) because various ``interesting'' scalar products may exist (see the discussion in \cite{CST}). In the present case, however, the situation is very simple: the only scalar product of interest on $\mathcal B$, and on its dual,  is induced by the
hermitian structure determined by the choice of the orthonormal basis $\{0,1,l,r\}$ on the vector space $\mathcal H$, and the two basis of matrix units (for $\circ$ and for $\, {\widehat \circ} \,$) are both orthonormal (a special feature of the $A_2$ case).
At the level of $\mathcal B$,  we can therefore identify the basis $\{\widehat{00}, \widehat{01}, \widehat{10}, \widehat{11}, \widehat{rr}, \widehat{rl}, \widehat{lr}, \widehat{ll}\}$  with $\{{00}, {01}, {10}, {11}, {rr}, {rl}, {lr}, {ll}\}$.
An identification like $rr = \widehat{rr}$ reads pictorially as follows:
$$
\setlength{\unitlength}{.1in}
\vertgraph {0}{1}{0}{1}{ 1}{0}{.5}\punit2
= \qquad
\setlength{\unitlength}{.1in}
\horgraph {0}{1}{0}{1}{ 0}{0}{.5}\punit2
$$

The obtained algebra $(\mathcal B, \star)$ is, by definition, isomorphic with the algebra $(\widehat{\mathcal B}, \, {\widehat \circ} \,)$, the  isomorphism $F$ being an analogue of the Fourier transform ($F(u) \star F(v) = F(u \, {\widehat \circ} \, v)$). Remember that the correspondance between $(\mathcal B, \Delta)$ and $(\widehat{\mathcal B}, \, {\widehat \circ} \,)$ does not involve any choice of a scalar product (or any star operation), but the correspondance between  $(\widehat{\mathcal B}, \, {\widehat \circ} \,)$ and  $(\mathcal B, \star)$ does. For this reason, we prefer to keep distinct notations (and we shall mostly use $\, {\widehat \circ} \,$ to remember that, in general, one does not need to specify any scalar product).

\paragraph{Note.} In general, once a basis of matrix units $e_I$ has been  fixed  for the multiplication $\circ$ in  $\mathcal B$ and once a basis of matrix units $E^A$ has been fixed for the  multiplication $\, {\widehat \circ} \,$ in $\widehat{\mathcal B}$, one 
may consider not only their respective dual basis $e^I$ and $E_A$, but also, if some scalar product has been 
chosen, the four basis $e_I^\sharp$ (in $\widehat{\mathcal B}$),  $e^I_\flat $ (in $\mathcal B$), $E_A^\sharp$ (in $\widehat{\mathcal B}$),  $E^A_\flat$ (in $\mathcal B$) resulting from the application of the 
the musical isomorphisms $\sharp$ and $\flat$ associated with the scalar product (the usual morphisms between a vector space and its dual). Such possibilities can be forgotten in the present case where the situation is very simple.

\subsubsection{The matrix realizations of the $\circ$ and $\star$ products of ${\mathcal B}$}
\label{sec:matrixrealizations}

We call $\circ$-realization, the (faithful) representation of $({\mathcal 
B},\circ)$  by matrices;  in particular, $\one$ should be the unit matrix.
We call $\star$-realization, the (faithful) representation of $({\mathcal 
B},\star)$, or of $( \widehat{\mathcal B}, \, {\widehat \circ} \,)$,  by matrices; in particular $\widehat \one$ should be the unit matrix. As we saw already, the basis elements $\{00,01,10,11,rr,rl,lr,ll\}$ 
are matrix units for these two algebras\footnote{This is not so for more complicated diagrams, not even for $A_n$, as soon as $n>2$.}  -- but not in the same way .  We summarize the results as
 follows (notation should be obvious).

$$
{\mathcal B} =_{\circ} \left(\begin{array}{cc}  
\left( \begin{array}{cc} 00 & 01 \\ 10 & 11 \end{array} \right) & {} \\ {} & 
\left( \begin{array}{cc} rr & rl \\ lr & ll \end{array} \right) \end{array} 
\right)  
$$

$$
{\mathcal B} =_{\star} \left(\begin{array}{cc}  \left( \begin{array}{cc} 00 & 
rr \\ ll & 11 \end{array} \right) & {} \\ {} & \left( \begin{array}{cc} 
01 & rl \\ lr & 10 \end{array} \right) \end{array} \right)
$$

In a sense, the above two equations summarize the present paper in a compact way.

\subsection{Writing the $\star$ product with matrices implementing 
the $\circ$ product}

Exercise: express the ``convolution  product'' $ \star$ in terms of  matrices that
are adapted to  the ``composition product''  $\circ$ of  $\mathcal B$.  To solve this problem, 
we take two  arbitrary element $u_1$ and $u_2$, calculate their product  $u_{1}\star u_{2}$ using the $\star$ multiplication table (or the $\, {\widehat \circ} \,$ table, which is the same) and rewrite the result in terms of matrices. 
\begin{eqnarray*}
u_{1} &=& (a_{1} 00 + b_{1} 01 + c_{1 } 10 + d_{1} 11 + \alpha_{1} 
rr + \beta_{1} rl + \gamma_{1} lr + \delta_{1} ll)\\
u_{2} &=& (a_{2} 00 + b_{2} 02 + c_{2 } 20 + d_{2} 22 + \alpha_{2} 
rr + \beta_{2} rl + \gamma_{2} lr + \delta_{2} ll)
\end{eqnarray*}
The result is as follows:
\begin{center}
{\small
$
\begin{array}{c c c c }
 \left(
\begin{array}{c|c}
    \begin{array}{cc}
    a_{1} & b_{1}  \\
    c_{1} & d_{1}
\end{array} & 0  \\ \hline
    0 & \begin{array}{cc}
    \alpha_{1} & \beta_{1}  \\
    \gamma_{1} & \delta_{1}
\end{array}
\end{array}
\right)
&\star&

 \left(
\begin{array}{c|c}
    \begin{array}{cc}
    a_{2} & b_{2}  \\
    c_{2} & d_{2}
\end{array} & 0  \\  \hline
    0 & \begin{array}{cc}
    \alpha_{2} & \beta_{2}  \\
    \gamma_{2} & \delta_{2}
\end{array}
\end{array}
\right)
&=

\end{array}
$
}
\end{center}
\begin{center}
{\small
$
 \left(
\begin{array}{c|c}
    \begin{array}{cc}
    a_{1}a_{2} + \alpha_{1}\delta_{2} & b_{1} b_{2} + \beta_{1} \gamma_{2}  \\
    c_{1} c_{2} + \gamma_{1} \beta_{2} & d_{1} d_{2} + \delta_{1} \alpha_{2}
\end{array} & 0  \\  \hline
    0 & \begin{array}{c|c}
    a_{1}\alpha_{2} + \alpha_{1} d_{2} & b_{1} \beta_{2} + \beta_{1} c_{2}  \\
   c_{1}\gamma_{2} + \gamma_{1} b_{2} &d_{1} \delta_{2} + \delta_{1} a_{2}
\end{array}
\end{array}
\right)
$
}
\end{center}

\section{Quantum groupo\"\i d  considerations}

\subsection{General weak Hopf algebras properties}

\subsubsection{Counit}

In the case of weak Hopf algebras,  the counit is not an homomorphism and the compatibility axiom
is modified as \cite{BohmSzlachanyi}:
\begin{eqnarray*}
\epsilon(xy) &=& \epsilon (x \one_{1}) \epsilon(\one_{2} y)
\end{eqnarray*}
These two equations can be
checked here, by using the  expression
of $\epsilon$ given previously.

\subsubsection{Star(s) and antipode}

A priori one may introduce several kinds of  natural conjugations -- besides complex conjugation -- on 
the {\sl vertical} diffusion graphs (of section \ref{sec:diffusiongraphs}).
\begin{itemize}
        \item One may flip globally the top and bottom labels.
    This operation $u \mapsto u^\dag$ corresponds  to the adjoint operation in the matrix realization of $(\mathcal B, \circ)$
       \item One may flip globally the left and right  labels.
    This operation $u \mapsto u^{\hat \dag}$ corresponds  to the adjoint operation in the matrix realization of $(\mathcal B, \star)$
    \item One may flip simultaneously  top and bottom labels, as well 
    as left 
    and right labels.  This will corresponds to  the antipode $S$.
    \end{itemize}

Operation $\dag$:
$ 00 \mapsto 00, 01 \mapsto 10 , 10 \mapsto 01 , 11 \mapsto 11 , rr 
\mapsto rr , rl \mapsto lr, 
lr \mapsto rl , ll \mapsto ll $

Operation $\hat \dag$:
$ 00 \mapsto 00, 01 \mapsto 01, 10 \mapsto 10, 11 \mapsto  11, rr \mapsto 
ll, rl \mapsto lr, 
lr \mapsto rl, ll \mapsto  rr$

Operation $S$:
$ 00 \mapsto 00, 01 \mapsto 10, 10 \mapsto 01, 11 \mapsto 11 , rr \mapsto 
ll, rl \mapsto rl, 
lr \mapsto lr , ll \mapsto rr $

In the first two cases, one should in principle also take the complex conjugate of the above expressions,  but this is  unnecessary (with the chosen basis) since the matrix units are real.

In the present case, and more generally for weak Hopf algebras, the compatibility axiom for 
antipode is modified as follows (\cite{BohmSzlachanyi}):
$$
S(x_{1}) x_{2} \otimes x_{3} = \one_{1} \otimes x \one_{2}
$$

\subsection{Fusion algebra and algebra of quantum symmetries}

Now we would like to multiply together  the projectors $\pi$ with the multiplication $\, {\widehat \circ} \,$ and 
multiply together the projectors $\varpi$ with the multiplication $\circ$.
This does not make sense, {\it a priori \/},  unless we replace the  elements of  the basis of $\mathcal B$ (or of $\widehat{\mathcal B}$) entering the expression of $\pi$ (or of $\varpi$)  by the corresponding elements belonging to the dual basis. In the case of the diagram $A_2$, we may forget about this difficulty.
 
\subsubsection{Multiplication table of the projectors $\pi_{i}$ : the fusion algebra $A(A_2)$ }

We already checked that the $\pi_i$ constitute a system of orthogonal and complete
minimal central projectors for $\circ$:
$ \pi_{0} \circ \pi_{0} = \pi_{0}$, 
$ \pi_{1} \circ \pi_{1} = \pi_{1}$,
$ \pi_{0} \circ \pi_{1} =  \0 $,
$ \pi_{0} + \pi_{1} = \one $.
These operators  generate linearly  the center of the $(\mathcal B, \circ)$.
We now discover that this space is closed for the other multiplication by calculating  the product of these elements for  the $\, {\widehat \circ} \,$ multiplication.
\begin{center}
$ \pi_{0} \, {\widehat \circ} \, \pi_{0} = \pi_{0}$, \,
$ \pi_{0} \, {\widehat \circ} \, \pi_{1} = \pi_{1}$,\,
$ \pi_{1} \, {\widehat \circ} \, \pi_{0} =   \pi_{1} $,\,
$ \pi_{1} \, {\widehat \circ} \, \pi_{1} = \pi_{0}$.\,
\end{center}
The  associative algebra spanned by the $\pi_{i}$ is therefore isomorphic with the graph algebra of 
$A_{2}$. This is an algebra with one generator ($\pi_1$) and the graph itself (or its adjacency matrix) encodes multiplication by the generator. $\pi_0$ is the unit . More generally, if we start from a diagram $G$ (of type $ADE$, to simplify),  the algebra obtained in this way is called the fusion algebra $A(G)$ of $G$ (it is always a graph algebra of type $A_n$).

\subsubsection{Multiplication table of the projectors $\varpi_{x}$ : the algebra of quantum symmetries $Oc(A_2)$}

We already checked that the $\varpi$ constitute a system of orthogonal and complete
minimal central projectors for $\, {\widehat \circ} \,$:
$ \varpi_{0} \, {\widehat \circ} \, \varpi_{0} = \varpi_{0}$, 
$ \varpi_{1} \, {\widehat \circ} \, \varpi_{1} = \varpi_{1}$,
$ \varpi_{0} \, {\widehat \circ} \, \varpi_{1} =  \0 $,
$ \varpi_{0} + \varpi_{1} = \widehat \one $.
These operators  generate linearly  the center of the $(\widehat{\mathcal B}, \, {\widehat \circ} \,)$.
We now discover that this space is closed for the other multiplication by calculating  the product of these elements for  the $\circ$ multiplication.
\begin{center}
$ \varpi_{0} \circ \varpi_{0} = \varpi_{0}$, \,
$ \varpi_{0} \circ \varpi_{1} = \varpi_{1}$,\,
$ \varpi_{1} \circ \varpi_{0} =   \varpi_{1} $,\,
$ \varpi_{1} \circ \varpi_{1} = \varpi_{0}$.
\end{center}
The  associative algebra spanned by the $\varpi_{i}$ is again isomorphic with the graph algebra of 
$A_{2}$.  More generally, if we start from a diagram $G$  the algebra obtained in this way is called the algebra of quantum symmetries $Oc(G)$ of $G$. The multiplication table of its generators is encoded by a graph called the Ocneanu graph of $G$. In the $A_{2}$ case, the original graph and its 
Ocneanu graph therefore coincide (more generally this is so when $G$ is of type $A_n$).


\subsection{Left and right counital subalgebras}

As shown in references \cite{BohmSzlachanyi} and \cite{NikshychVainerman}, the so called left and right counital subalgebras play an
important role in the general theory of weak Hopf algebras. In this section we  shall illustrate four distinct ways to determine them and check explicitly, in this particular case,  several general properties.

\subsubsection{Using the coproduct of the unit}

We already calculated $\Delta \one$ in section \ref{sec:deltaone}. The result suggests to set
\begin{eqnarray*}
\{e^{0}, e^{1}\} &\doteq &\{00 + rr, ll + 11\}
\\
\{e_{0}, e_{1}\} &\doteq & \{00 + ll, rr + 11\}
\end{eqnarray*}
so that
$$
\Delta \one = e_{i} \otimes e^{i}
$$
We call ${\mathcal B}_{s}$ the linear span of $e^{i}$ (source
counital subalgebra) and ${\mathcal B}_{t}$ the linear span of $e_{i}$ (target
counital subalgebra).
Call $s$ (source) and $t$ (target) the maps that associate to an elementary path its initial or final vertex.
For $A_2$, we have\footnote{The ``bold'' $\bf 0$ denotes zero or  the null vector whereas $0$ denotes the first vertex of $A_2$.} $s(0)=s(r)= 0$, $s(l)=s(1) = 1$, and $t(0)=t(l)= 0$, $t(r)=t(1)=1$.
$$
dim({\mathcal B}_{s}) = dim({\mathcal B}_{t}) = \epsilon(\one) = 1 + 1 + \bf 0 + \bf 0 = 2 = 
\#\{\mbox{vertices of  $A_{2}$}\}
$$
As we shall see later, this definition of $e_{i}$ and $e^{i}$ can
also be obtained from a general characterization of
target and source coideal subalgebras.

\subsubsection{Using the coproduct of the unit (second method)}

The simplest way to determine ${\mathcal B}_{s} $ is probably to use the property
$$
{\mathcal B}_{s} = \{ <u \otimes id, \Delta \one>, u \in \hat {\mathcal B} \} = \{
00+rr,11+ll \} 
$$
In the same way, we determine
$$
{\mathcal B}_{t} = \{ <id \otimes u, \Delta \one>,  u \in \hat {\mathcal B}  \} 
=\{00+ll,11 + rr \} 
$$

\subsubsection{Source and target counital maps}

We can define these maps  in two equivalent ways\footnote{ $\Delta h = h_1 \otimes h_2$ stands for 
the coproduct of an arbitrary element $h$ (a summation is understood).}:
\begin{eqnarray*}
\epsilon_{t}(h) &=& (id \otimes \epsilon)((\one \otimes h) \circ  \Delta(\one)) \\
\epsilon_{s}(h) &=& (\epsilon \otimes id )(\Delta(\one) \circ (h \otimes \one) )
\end{eqnarray*}
\begin{eqnarray*}
\epsilon_{t}(h) &=& S(h_{1}) \circ h_{2} \\
\epsilon_{s}(h) &=& h_{1} \circ S(h_{2})
\end{eqnarray*}

Using the first definition (with the counit), we determine $\epsilon_{s}$:
$$
\epsilon_{s}\left\{
\begin{array}{c}
00 \\ 01 \\ 10 \\ 11 \\ rr \\ rl \\ lr \\ ll
\end{array}
\right\}
= \left\{
\begin{array}{c}
\epsilon{(00)} \otimes (00 + rr), \\
\epsilon{(01)} \otimes (00 + rr), \\
\epsilon{(10)} \otimes (ll + 11), \\
\epsilon{(11)} \otimes (ll + 11) ,\\
\epsilon{(rr)} \otimes (ll + 11), \\
\epsilon{(rl)} \otimes (ll + 11), \\
\epsilon{(lr)} \otimes (00 + rr), \\
\epsilon{(ll) } \otimes (00 + rr) 
\end{array}
\right\}
=
\left\{
\begin{array}{c}
    00 + rr \\
    00 + rr \\
    ll + 11 \\
    ll + 11 \\
    \bf 0 \\
     \bf 0 \\
      \bf 0 \\
       \bf 0 
 \end{array}  
 \right\}
$$

The image of the source couinital map $\epsilon_s$ is ${\mathcal B}_{s}$, it is the span 
of  $ \{00+rr,11+ll \}$, as expected. The image of the target couinital map $\epsilon_t$ is ${\mathcal B}_{t}$, it is the span of  $ \{00+ll,11+rr \}$.

Another way to compute the images of $\epsilon_{s}$ and 
$\epsilon_{t}$ is to use their determination using the antipode ( 
this provides a check that our definition of the antipode is correct).
Examples:
\begin{eqnarray*}
\epsilon_{s}(00) &=& 00 \circ S(00) + rr \circ S(ll) = 00 \circ 00 + rr
\circ rr = 00 + rr \\
\epsilon_{s}(01) &=& 01 \circ S(01) + rl \circ S(lr) = 01 \circ 10 + rl
\circ lr = 00 + rr \\
\epsilon_{s}(10) &=& 10 \circ S(10) + lr \circ S(rl) = 10 \circ 01 + lr
\circ rl = 11 + ll \\
\epsilon_{s}(11) &=& 11 \circ S(11) + ll \circ S(rr) = 11 \circ 11 + ll
\circ ll = 11 + ll \\
\end{eqnarray*}

\subsubsection{General properties}
\paragraph {From ${\mathcal B}_t$ to ${\mathcal B}_s$.}

We can check that ${\mathcal B}_t$ and ${\mathcal B}_s$ commute, that 
 $S$ maps ${\mathcal B}_t$ to ${\mathcal B}_s$ (and conversely),  and that  $S(\epsilon_{t}(h)) = \epsilon_{s}(S(h))$.

\paragraph {Characterization of ${\mathcal B}_{s}$.}
We can also check the following property $$ \forall h \in {\mathcal B}_{s}, \Delta (h) = (1_{(1)} \circ h )
\otimes 1_{(2)}$$ For instance
$\Delta (00 + rr) = 00 \otimes 00 + rr \otimes ll + 00 \otimes rr + 
rr \otimes 11 = 1_{(1)} \circ (00 + rr) \otimes 1_{(2)}.$

\paragraph {Subalgebra and (co)-ideal properties.}
Multiplication rules in ${\mathcal B}_{s}$ for the $\circ$ multiplication read
$e^{0} \circ e^{0} = e^{0}$, $e^{1}\circ e^{1}=e^{1}$, $e^{0}\circ e^{1}=e^{1}\circ e^{0}=\bf 0$.
In the case of $\, {\widehat \circ} \,$ multiplication we find:
\begin{eqnarray*}
\{00,01,10,11,rr,rl,lr,ll\} \, {\widehat \circ} \, e^{0} &=& \{e^{0}, {\bf 0 , \bf 0, \bf 0, \bf 
0, \bf 0 ,\bf 0 }, e^{1}\} 
\\
\{00,01,10,11,rr,rl,lr,ll\} \, {\widehat \circ} \, e^{1} &=& \{ {\bf 0 , \bf 0, \bf 0}, 
e^{1}, e^{0}, {\bf 0, \bf 0 ,\bf 0} \}
\end{eqnarray*}
However
\begin{eqnarray*}
e^{0} \, {\widehat \circ} \, \{00,01,10,11,rr,rl,lr,ll\} &=& \{00, \bf  0, \bf  0,rr, rr,
\bf 0, \bf  0, 00\}
\\
e^{1} \, {\widehat \circ} \, \{00,01,10,11,rr,rl,lr,ll\} &=&\{ll, \bf  0, \bf  0,11,11, \bf  
0, \bf  0,ll\}
\end{eqnarray*}
Notice that $e^0 \circ  rl = 00 \circ rl + rr \circ rl = rl $ which does not belong to ${\mathcal B}_{s}$.

We see that  ${\mathcal B}_{s}$  is indeed a subalgebra for $\circ$, but not an ideal for this multiplication.
It is a right ideal for $\, {\widehat \circ} \,$, but not a bilateral ideal.
Conclusion (rephrased in terms of coproduct): ${\mathcal B}_{s}$ is a coideal subalgebra of ${\mathcal B}$
We have a similar conclusion for ${\mathcal B}_{t}$.

\paragraph {Separability.}
The separability element $e_s \in {\mathcal B}_s
\otimes {\mathcal B}_s$ is such that $m(e_s)=\one$, $(a \otimes \one) \circ e_s = e_s \circ 
(\one \otimes a)$, $(\one \otimes a) \circ e_s = e_s \circ 
(a \otimes \one)$ for all $a \in {\mathcal B}_s$,  with similar properties for the element $e_t \in  {\mathcal B}_t
\otimes {\mathcal B}_t$. The separability elements are:
$$
e_{s} =  (S \otimes id) \Delta \one  = (00 + rr)\otimes (00 + 
rr) + (ll + 11) \otimes (ll + 11) = e^0 \otimes e^0 + e^1 \otimes e^1
$$
$$
e_{t} = (id \otimes S) \Delta \one =  (00 + ll) \otimes (00 + ll) + (rr + 11) \otimes (rr + 11) = 
e_0 \otimes e_0 + e_1 \otimes e_1
$$

\paragraph{A special bilinear form on the source and target subalgebras.}
One could think of defining a bilinear form on ${\mathcal {B}}$ by $\rho, \tau  \mapsto \epsilon(a^\dag b)$.
However, we remember that $\epsilon$ in our case is not an algebra
homomorphism, and it happens that only the restriction of the above bilinear form to the 
source and target subalgebra coideals  defines an interesting scalar product \cite{BohmSzlachanyi}.
For instance,  the basis $e^{i}$ is  orthonormal for this scalar 
product in ${\mathcal B}_{s}$.  Let us ckeck this explicitly:
$\epsilon( (00+rr)^\dag \circ (00 + rr)) = 
\epsilon( (00+rr) \circ (00 + rr)) = \epsilon (00) = 1$.
This ``scalar product'' is not the one induced by the (genuine) scalar 
product on essential paths (which would give, in 
the above example,  a numerical result equal to  $2$ rather than equal to $1$). 

We can also check (use the definition of  antipode) that if
$e^{i} = \{00+rr,11+ll \} $ is an orthonormal basis of ${\mathcal B}_{s}$, so that 
$(e^{i})^\dag = \{00+rr,11+ll \} $,  then 
$e_{i} = S((e^{i})^\dag) = \{00+ll,11+rr \} $ is an orthonormal basis of 
${\mathcal B}_{t}$.

\subsection{Representation theory}

In the category of quantum groupo\"\i ds,
the tensor product of representations $V$ and $W$ is {\bf not} $V  \otimes 
W$ but\footnote{We do not write explicitly the module action maps}
 $$V \widetilde \otimes W \doteq \Delta(1) \circ  (V \otimes W)$$

The unit object for this (modified) tensor product is not the usual 
``trivial'' representation, but the coideal subalgebra ${\mathcal B}_{s}$.
So the trivial representation, here, is $2$ dimensional.
Let us check this. First of all the action of ${\mathcal B}$ on ${\mathcal B}_{s}$ is 
defined by $\rho \in {\mathcal B} \triangleright u \in {\mathcal B}_{s} = 
\epsilon_{s}(\rho u)$.
If we represent $f$ by a direct sum of two block matrices $(2,2)$ 
equal to {\small $ \left(\begin{array}{cc}  a & b \\ c & d  \end{array} \right)\oplus 
\left( \begin{array}{cc} e & f\\ g & h  \end{array}\right) $} and 
$u$ as the diagonal matrix $diag((\alpha, \beta), (\alpha, \beta))$, 
we compute $\rho \triangleright u$ and find $diag((a \alpha + b\beta, c 
\alpha + d \beta), (a \alpha + b\beta, c 
\alpha + d \beta))$. So this action is equivalent to the 
two-dimensional
representation that we called $\pi_{0}$ (projector corresponding to 
the first block) but we already knew that it 
acted as the unit because of the $\, {\widehat \circ} \,$ multiplication rules of the $\pi_{j}$.
In other words the unit object for the tensor product, in this 
category, is a two dimensional representation  equivalent 
to $\pi_{0}$. Therefore -- and
fortunately --   the unit object is irreducible  (this is not necessarily so for arbitrary weak Hopf 
algebras\ldots)

\section{Remarks}

\subsection{An associative algebra structure on $\mathcal H$ ?}
\label{sec:concat}
Let us compare  $$rl \circ lr = rr   \qquad {\mbox and} \qquad  rl \star lr = 01.$$
\ie
$$
\setlength{\unitlength}{.1in}
\vertgraph {0}{1}{1}{0}{ 1}{0}{.5}\punit2
\circ
\setlength{\unitlength}{.1in}
\vertgraph {1}{0}{0}{1}{ 1}{0}{.5}\punit2
=
\setlength{\unitlength}{.1in}
\vertgraph {0}{1}{0}{1}{ 1}{0}{.5}\punit2
$$
$$
\setlength{\unitlength}{.1in}
\vertgraph {0}{1}{1}{0}{ 1}{0}{.5}\punit2
\star
\setlength{\unitlength}{.1in}
\vertgraph {1}{0}{0}{1}{ 1}{0}{.5}\punit2
=
\setlength{\unitlength}{.1in}
\vertgraph {0}{0}{1}{1}{ 0}{0}{.5}\punit2
$$
This result suggests that, when expressed in terms of vertical diffusion graphs (matrix units for endomorphisms of essential paths), the result of a $\star$  multiplication  is obtained by concatenating seperately the top and bottom paths. This observation is at the root of the following idea: try to define
a product (call it $\times$) in the vector space of essential paths { $\mathcal H$}, such that $\star$ appears as the tensor square of $\times$. It is not too difficult to guess that the following multiplication table (in $\mathcal H$) works:
$$
\begin{array}{c|cccc}
\times \nearrow &    0 & r & 1 & l \\
    \hline \\
0 & 0 & r & . &.  \\
r & . & . & r & 0  \\
1 & . & .& 1 & l  \\
l & l & 1 & . & . \\
\end{array}
$$
This product is associative (it is enough to check the cubics). If we now compare this product  with the multiplication $\star$ in $\mathcal B$, it is a simple matters to check that, indeed (we restore tensor product signs), 
$$ (u \otimes v) \star (u'  \otimes  v')  \doteq (u \times u')  \otimes  (v \times v') $$
The purpose of this observation is to suggest  possible generalizations that would be valid for more general graphs. However this is not  straightforward. First of all, the concatenation of two essential paths is not essential in general, so that one has anyway to reproject from $Paths$ to $EssPaths$. Moreover
there is already a natural product for the product of two paths, called concatenation product, and defined as concatenation of the two paths  if they match and zero otherwise,  but the previous operation $\times$ is different. For example, the concatenation of $r$ and $l$ is the path of length $2$ equal to $0 \rightarrow 1 \rightarrow 0$ which is not essential and whose projection on essential paths vanishes, but  if we remove the non-essential part
(the  round trip) around the junction point (the vertex $v_1$) we are left with the vertex $v_0$, which is indeed the value of $r \times l$.
 The $\times$ multiplication in $\mathcal H$ appears therefore as concatenation followed by  pairwise removal of 
non essential pieces around the junction point. The above ansatz  is precise enough to recover the definition of $\times$ given for the diagram $A_{2}$ and one expects that such a product, in more general cases, would be filtered,  not graded (the multiplication of two essential paths of length $p$ and $q$ would incorporate not only paths of length $p+q$ but also paths of length $p+q-2$, $p+q-4$, \etc, with appropriate coefficients). A precise definition of $\times$,  for more general diagrams, even for those belonging to the $A$ series, is however not available and  we do not know -- with the exception  of $A_2$-- how to obtain the $\star$ product on $\mathcal B$ as the tensor square  of any associative multiplication  (possibly re-projected) defined in the space of  essential paths.
For this reason, the explicit definition of the multiplications  $\, {\widehat \circ} \,$ (or $\star$)  usually requires the 
calculation of all Ocneanu cells.

\subsection{Generalizations}

 We have explicitly constructed the weak Hopf algebra $\mathcal B$ associated with the choice of the $A_2$ diagram.  For diagrams belonging to the $A_n$ family, the discussion is similar:  the algebra structures on  $\mathcal B$ and on its dual are  both isomorphic to the same direct  sum of full matrix algebras; the associated fusion algebra and associated algebra of quantum symmetries are also both isomorphic to the graph algebra of the given diagram $A_n$.  A construction of  $\mathcal B$ is explicitly given, for several examples of this type, in reference \cite{CST}.
For diagrams of  type $D$ or $E$, the situation is more involved: the two algebra structures on $\mathcal B$ and its dual are different. If the Coxeter number of the diagram is $\kappa$, the first algebra structure (the composition law $\circ$) is given by a direct sum of  $\kappa -1$ blocks, and the fusion algebra is isomorphic with the graph algebra of $A_{\kappa -1}$. The algebra structure for the convolution law $\star$ is also semi-simple, but the number of blocks, their dimensions, and the structure of the algebra of quantum symmetries has to be determined in  each case. The determination of  fusion algebras and algebra of quantum symmetries, for general diagrams,  can actually be done without having to construct explicitly the  bialgebra $\mathcal B$ (\ie without determining the two sets of matrix units or the corresponding cells). We refer to the articles quoted in the introduction for a detailed study along these lines.

The notation used to denote matrix units (diffusion graphs) or the dual notation (using double triangles) is sufficient\footnote{One should be cautious since the same notation, for other authors,  could involve proportionality factors.} here, but a study of more general cases requires the introduction of several types of lines. Indeed,  the labels for the  different blocks of the two associative structures are in one-to-one correspondance with generators of the fusion algebra or with those of the algebra of quantum symmetries, but for $A_n$ diagrams, both types of generators are also in one-to-one correspondance with the vertices of the original diagram. For more general cases, these objects are distinct and one has to introduce three types of lines ({or dually, two types of colors for the vertices used in the diagrammatic description using double triangles}): one (say $a$) for vertices of the given diagram, one (say $n$) for vertices of the corresponding fusion graph and one (say $x$) for vertices of the corresponding Ocneanu graph.

As mentioned in the introduction, the theory that we illustrated on the very simple example of the $A_2$ diagram can be generalized to all $ADE$ diagrams and can even be further generalized to members of higher Coxeter-Dynkin systems.


\section{Acknowledgements}

The present understanding of the author concerning the structures illustrated here results from many discussions  with friends whom I want to thank here :
Ariel Garcia, Marina Huerta, Hugo Montani, Oleg Ogievetsky, Gil Schieber, Roberto Trinchero, and Adrian Ocneanu.



\end{document}